\documentclass{webofc}
\usepackage[varg]{txfonts}   
\begin{document}
\title{Factorization, resummation and sum rules for heavy-to-light form factors}
%
%

\author{ Yu-Ming Wang \thanks{\email{yu-ming.wang@univie.ac.at}}
}

\institute{Fakult\"{a}t f\"{u}r Physik, Universit\"{a}t Wien, Boltzmanngasse 5, 1090 Vienna, Austria
\and
School of Physics, Nankai University, 300071 Tianjin, China
          }

\abstract{%
Precision calculations of heavy-to-light form factors are essential to sharpen our understanding
towards the strong interaction dynamics of the heavy-quark system and to shed light on
a coherent solution  of flavor anomalies. We briefly review factorization  properties of heavy-to-light form factors
in the framework of QCD factorization in the heavy quark limit and discuss the recent progress
on the QCD calculation of $B \to \pi$ form factors from the light-cone sum rules with the $B$-meson distribution amplitudes.
Demonstration of QCD factorization for the vacuum-to-$B$-meson correlation function used in the sum-rule construction and
resummation of large logarithms in the short-distance functions entering the factorization theorem are presented in detail.
Phenomenological implications  of the newly derived sum rules for $B \to \pi$ form factors are further addressed with a particular
attention to the  extraction of the CKM matrix element $|V_{ub}|$.
}
\maketitle
\section{Introduction}

Heavy-to-light form factors serve as fundamental inputs of describing many exclusive heavy hadron decays
which are of great phenomenological interest to the ongoing and forthcoming collider experiments.
Extensive efforts have been devoted to develop systematic theoretical frameworks for the precision calculation
of heavy-to-light form factors in QCD (see, for instance\cite{Beneke:2000wa,Ball:2004ye,Duplancic:2008ix,Khodjamirian:2011ub,Wang:2008sm,Li:2012nk,Lu:2009cm}).
In addition to the nonperturbative hadronic distribution amplitudes (DA), the symmetry-conserving ``soft" form factors
have to be introduced in QCD factorization for  $B \to \pi$ form factors due to the emergence of rapidity divergences
in the corresponding convolution integrals. An alternative approach to compute $B \to \pi$ form factors is QCD light-cone
sum rules (LCSR) constructed from the vacuum-to-$B$-meson correlation functions with the aid of the dispersion relations and
parton-hadron duality approximation \cite{Khodjamirian:2005ea,Khodjamirian:2006st,DeFazio:2005dx,DeFazio:2007hw}.
In the following, we will first establish QCD factorization for the vacuum-to-$B$-meson correlation function at one loop employing
the method of regions and perform the resummation of large logarithms in the hard and jet functions with the renormalization-group
approach in section \ref{sec: factorization}, where the resummation improved LCSR for $B \to \pi$ form factors
and a new determination of the CKM matrix element $|V_{ub}|$ are also presented.

\section{The LCSR for the $B \to \pi$ form factors at ${\cal O}(\alpha_s)$}
\label{sec: factorization}

The construction of LCSR for the $B \to \pi$ form factors can be achieved with the following
correlation function \cite{Wang:2015vgv}:
\begin{eqnarray}
\Pi_{\mu}(n \cdot p,\bar n \cdot p)&=& \int d^4x ~e^{i p\cdot x}
\langle 0 |T\left\{\bar{d}(x) \not \! n \, \gamma_5 \, u(x),
\bar{u}(0) \, \gamma_\mu  \, b(0) \right\}|\bar B(p+q) \rangle
\nonumber\\
&=&\Pi(n \cdot p,\bar n \cdot p) \,  n_\mu +\widetilde{\Pi}(n \cdot p,\bar n \cdot p) \, \bar n_\mu\,,
\label{correlator: definition}
\end{eqnarray}
where $p+q=m_B \, v$ and the two light-cone vectors satisfy the relations $n \cdot v=\bar n \cdot v =1$.
To facilitate the perturbative calculation of short-distance functions in the factorization formula of
(\ref{correlator: definition}), we need to establish the power counting scheme for the external momentum $p_{\mu}$
\begin{eqnarray}
n \cdot p \simeq \frac{m_B^2+m_{\pi}^2-q^2}{m_B}= 2 E_{\pi}  \,,
\qquad \bar n \cdot p \sim \cal{O} ({\rm \Lambda_{QCD}}) \,.
\label{power counting scheme}
\end{eqnarray}
Applying the light-cone operator-produce-expansion (OPE) at space-like $p^2$ yields the tree-level factorization formula
\begin{eqnarray}
\widetilde{\Pi}(n \cdot p,\bar n \cdot p) &=& \tilde f_B(\mu) \, m_B \,
\int_0^{\infty} d \omega^{\prime} \,
\frac{\phi_B^-(\omega^{\prime})}{\omega^{\prime} - \bar n \cdot p- i \, 0}
+ {\cal O}(\alpha_s) \,, \nonumber \\
\Pi(n \cdot p,\bar n \cdot p) &=& {\cal O}(\alpha_s) \,.
\label{factorization of correlator:tree}
\end{eqnarray}
in the heavy quark limit, where the $B$-meson light-cone DA in the coordinate space is defined as
\begin{eqnarray}
\langle  0 |\bar d_{\beta} (\tau \, \bar{n}) \, [\tau \bar{n}, 0] \, b_{\alpha}(0)| \bar B(p+q)\rangle
= - \frac{i \tilde f_B(\mu) \, m_B}{4}  \bigg \{ \frac{1+ \! \not v}{2} \,
\left [ 2 \, \tilde{\phi}_{B}^{+}(\tau) + \left ( \tilde{\phi}_{B}^{-}(\tau)
-\tilde{\phi}_{B}^{+}(\tau)  \right )  \! \not n \right ] \, \gamma_5 \bigg \}_{\alpha \beta}\,.
\end{eqnarray}

Applying the standard definitions of $B \to \pi$ form factors and the pion decay constant
\begin{eqnarray}
\langle \pi(p)|  \bar u \gamma_{\mu} b| \bar B (p_B)\rangle
&=& f_{B \pi}^{+}(q^2) \, \left [ p_B + p -\frac{m_B^2-m_{\pi}^2}{q^2} q  \right ]_{\mu}
+  f_{B \pi}^{0}(q^2) \, \frac{m_B^2-m_{\pi}^2}{q^2} q_{\mu} \,, \nonumber \\
\langle \pi(p)  |\bar d \! \not n \, \gamma_5 \, u |  0 \rangle &=& - i \, n \cdot p \, f_{\pi} \,,
\end{eqnarray}
it is straightforward to derive the hadronic dispersion relation for the correlator (\ref{correlator: definition})
\begin{eqnarray}
&& \Pi_{\mu}(n \cdot p,\bar n \cdot p) \nonumber \\
&& = \frac{f_{\pi} \, n \cdot p \, m_B}{2 \, (m_{\pi}^2-p^2)}
\bigg \{  \bar n_{\mu} \, \left [ \frac{n \cdot p}{m_B} \, f_{B \pi}^{+} (q^2) + f_{B \pi}^{0} (q^2)  \right ]
+   n_{\mu} \, \frac{m_B}{n \cdot p-m_B}  \, \,
\left [ \frac{n \cdot p}{m_B} \, f_{B \pi}^{+} (q^2) -  f_{B \pi}^{0} (q^2)  \right ] \bigg \} \, \nonumber \\
&& \hspace{0.4 cm} + \int_{\omega_s}^{+\infty}  \, d \omega^{\prime} \, \frac{1}{\omega^{\prime} - \bar n \cdot p - i 0} \,
\left [ \rho^{h}(\omega^{\prime}, n \cdot p)  \, n_{\mu} \,
+ \tilde{\rho}^{h}(\omega^{\prime}, n \cdot p)  \, \bar{n}_{\mu}  \right ] \,,
\end{eqnarray}
where the threshold parameter $\omega_s$ in the pion channel scales as $\Lambda^2/m_b$.
Matching the hadronic and OPE representations of the vacuum-to-$B$-meson correlation function leads to
\begin{eqnarray}
f_{B \pi}^{+}(q^2) &=& \frac{\tilde f_B(\mu) \, m_B}{f_{\pi} \,n \cdot p} \, {\rm exp} \left[{m_{\pi}^2 \over n \cdot p \,\, \omega_M} \right]
\int_0^{\omega_s} \, d \omega^{\prime} \, e^{-\omega^{\prime} / \omega_M} \,  \phi_B^{-}(\omega^{\prime}) + {\cal O}(\alpha_s) \,,
\nonumber \\
f_{B \pi}^{0} (q^2) &=&  \frac{n \cdot p}{m_B} \, f_{B \pi}^{+} (q^2) +  {\cal O}(\alpha_s) \,.
\label{the form-factor relation}
\end{eqnarray}

\begin{figure}
\centering
\includegraphics[width=12 cm,clip]{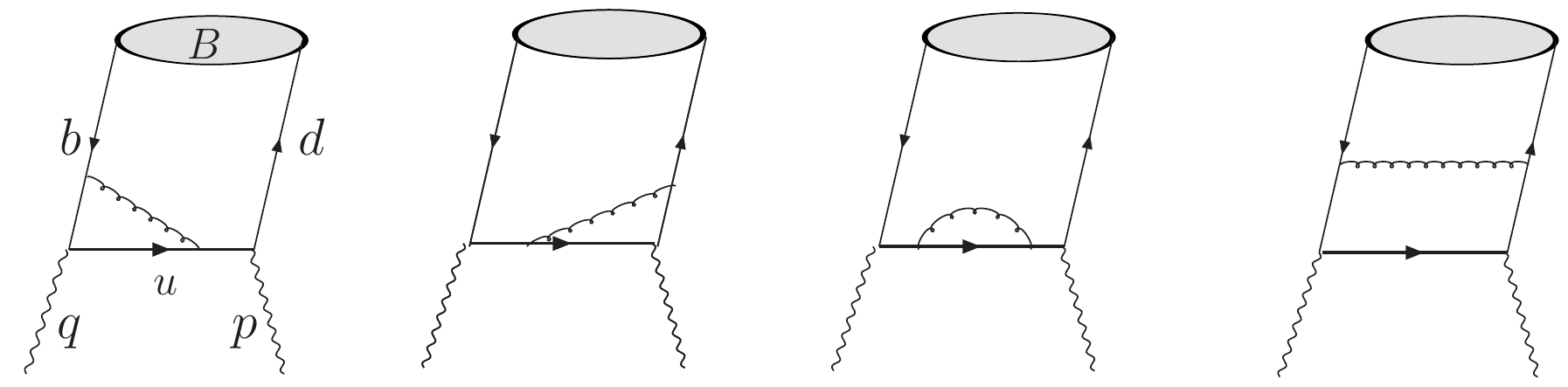} \\
\hspace{0.5 cm}(a) \hspace{2.5 cm} (b)\hspace{2.5 cm} (c) \hspace{3.0 cm} (d) \\
\vspace*{0.1cm}
\caption{Next-to-leading-order QCD correction to the correlation function $\Pi_{\mu}(n \cdot p,\bar n \cdot p)$. }
\label{fig: correlator at NLO}       
\end{figure}

\begin{figure}
\centering
\includegraphics[width=10 cm,clip]{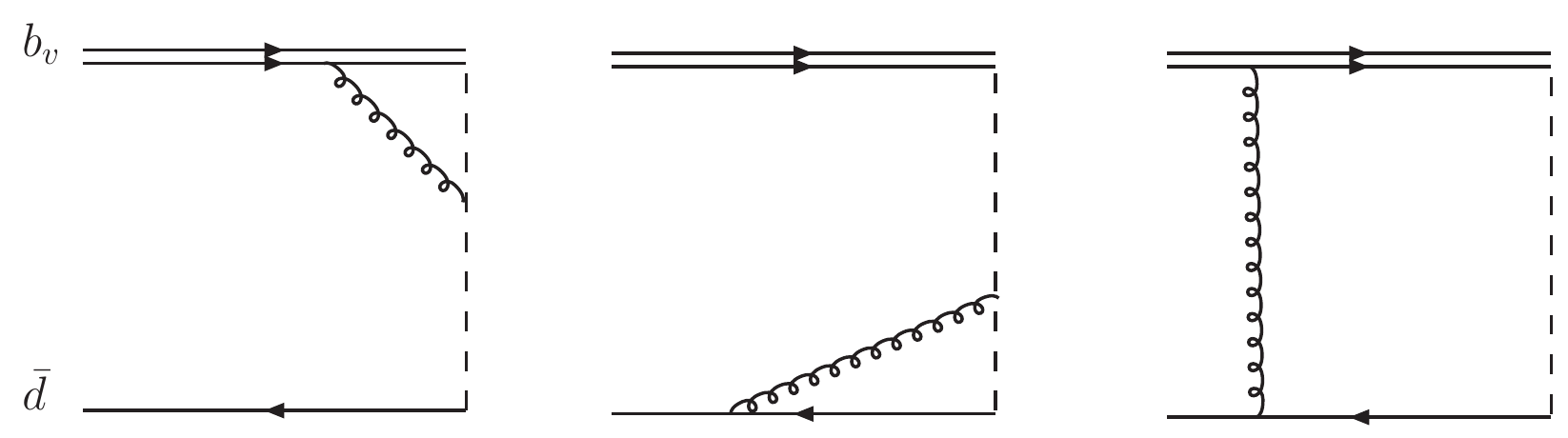} \\
\hspace{0.5 cm}(a) \hspace{3.5 cm} (b)\hspace{3.5 cm} (c)  \\
\vspace*{0.1cm}
\caption{One-loop effective diagrams for the infrared subtraction. }
\label{fig: IR subtraction}       
\end{figure}

Now we are in a position to demonstrate QCD factorization for $\Pi_{\mu}(n \cdot p,\bar n \cdot p)$ at one loop
using the diagrammatical approach. Technically, this amounts to justifying the equivalence between the soft contribution
to the one-loop QCD diagrams displayed in figure \ref{fig: correlator at NLO}
and the infrared subtraction shown in figure \ref{fig: IR subtraction} in the heavy quark limit and extracting the
hard and hard-collinear contributions to the correlation function at leading power in $\Lambda/m_b$.
We will apply the standard strategies to establish the diagrammatical factorization for QCD
Green functions at one-loop accuracy \cite{Wang:2015ndk,Wang:2016qii}:
\begin{itemize}
\item{Identify the leading regions of the QCD amplitudes with the power counting scheme (\ref{power counting scheme});}
\item {Evaluate the loop integrals with the method of regions and extract the ``bare" perturbative kernels;}
\item{Apply the ultraviolet renormalization and perform the infrared subtraction; }
\item{Substitute the momentum-space light-cone projector of the $B$-meson.}
\end{itemize}

Taking the weak vertex diagram displayed in figure \ref{fig: correlator at NLO}(a) as an example,
the corresponding QCD amplitude can be written as \cite{Wang:2015vgv}
\begin{eqnarray}
\Pi_{\mu,  \, weak}^{(1)}
&=& \frac{g_s^2 \, C_F}{2 \, (\bar n \cdot p -\omega)} \,
\int \frac{d^D \, l}{(2 \pi)^D} \,   \frac{1}{[(p-k+l)^2 + i 0][(m_b v+l)^2 -m_b^2+ i 0] [l^2+i0]}  \nonumber  \\
&& \bar d(k)  \! \not n \,\gamma_5  \,\! \not {\bar n} \,\, \gamma_{\rho}  \, (\! \not p - \! \not k  + \! \not l)
\, \gamma_{\mu} \, (m_b  \! \not v +  \! \not l+ m_b )\, \gamma^{\rho} \, b(v)  \,.
\label{diagram a: expression}
\end{eqnarray}
Employing the power counting scheme (\ref{power counting scheme}) one can readily identify that the leading power contributions
to $\Pi_{\mu,  \, weak}^{(1)}$ come from the hard, hard-collinear and soft regions.
Expanding the QCD amplitude $\Pi_{\mu,  \, weak}^{(1)}$ in terms of the powers of $\Lambda/m_b$ in the soft region
and keeping the leading-order terms yield
\begin{eqnarray}
\Pi_{\mu,  \, weak}^{(1),s}
&=& \frac{g_s^2 \, C_F}{2 \,( \bar n \cdot p -\omega) } \,
\int \frac{d^D \, l}{(2 \pi)^D} \, \frac{1}{[\bar n \cdot (p-k+l) + i 0][v \cdot l + i 0] [l^2+i0]}  \nonumber  \\
&& \bar d(k) \,\, \! \not n \,\,   \gamma_5  \,\,  \! \not {\bar n} \,\, \gamma_{\mu}  \,\,  b(v)  \, \,.
\label{diagram a: soft subtraction}
\end{eqnarray}
On the other hand, the one-loop correction to the $B$-meson DA can be computed with the Wilson-line Feynman rules
\begin{eqnarray}
\Phi_{b \bar d, \, a}^{\alpha \beta\,, (1)} (\omega, \omega^{\prime})
&=& i \, g_s^2 \, C_F\, \int \frac{d^D \, l}{(2 \pi)^D} \,
\frac{1}{[\bar n \cdot l + i 0][v \cdot l + i 0] [l^2+i0]}  \nonumber  \\
&& \times [\delta(\omega^{\prime}-\omega-\bar n \cdot l)-\delta(\omega^{\prime}-\omega)] \,
[\bar d(k)]_{\alpha}  \, [b(v)]_{\beta} \,\,,
\label{effective diagram a: wave function}
\end{eqnarray}
from which the infrared subtraction term can be deduced as
$ \Phi_{b \bar d ,  \, a}^{(1)}  \otimes T^{(0)} = \Pi_{\mu,  \, weak}^{(1), \, s}$,
verifying QCD factorization for the vacuum-to-$B$-meson correlation function at one loop diagrammatically.
By proceeding in a similar way, the hard-colliner contribution from  the weak vertex diagram
can be extracted as follows:
\begin{eqnarray}
\Pi_{\mu, weak}^{(1), hc}&=& \frac{g_s^2 \, C_F}{2 (\bar n \cdot p -\omega)} \,
\int \frac{d^D \, l}{(2 \pi)^D} \, \frac{2 \, m_b \, n \cdot (p+l)}
{[ n \cdot (p+l) \, \bar n \cdot (p-k+l) + l_{\perp}^2  + i 0][ m_b \, n \cdot l+ i 0] [l^2+i0]} \nonumber  \\
&& \bar d(k) \,\, \! \not n \,\,   \gamma_5  \,\,  \! \not {\bar n} \,\, \gamma_{\mu}  \,\,  b(p_b) \,.
\label{diagram a: hard-collinear subtraction}
\end{eqnarray}
which can be further computed with the loop integrals collected in the Appendix A of \cite{Wang:2015vgv}.
Finally, expanding the QCD amplitude (\ref{diagram a: expression}) in the hard region gives rise to
\begin{eqnarray}
\Pi_{\mu,  \, weak}^{(1), \, h}&=& \frac{\alpha_s \, C_F}{4 \, \pi} \, \tilde f_B(\mu) \, m_B \, 
\frac{\phi_{b \, \bar d}^{-}(\omega)}{\bar n \cdot p -\omega} \, 
\bigg \{ \bar n_{\mu}  \bigg [ {1 \over \epsilon^2} +
{1 \over \epsilon} \, \left ( 2 \, \ln {\mu \over  n \cdot p} + 1  \right ) + 2 \, \ln^2 {\mu \over  n \cdot p}
 + 2 \,\ln {\mu \over  m_b} \nonumber \\
&& -\ln^2 r - 2 \, {\rm Li_2} \left (- {\bar r \over r} \right )  
+{2-r \over r-1} \, \ln r +{\pi^2 \over 12} + 3 \bigg ]
 + n_{\mu}  \, \left [ {1 \over r-1} \, \left ( 1 +  {r \over \bar r}  \, \ln r  \right ) \right ]  \,  \bigg \} \,,
\label{diagram a: result of the hard region expression}
\end{eqnarray}
with $r=n \cdot p/m_b$ and $\bar r = 1-r$.

Along the same vein,  one can evaluate the leading power contributions to the remaining diagrams
in figure \ref{fig: correlator at NLO} and  the resulting factorization formulae for the correlation function
are given by
\begin{eqnarray}
\Pi &=& \tilde{f}_B(\mu) \, m_B \sum \limits_{k=\pm} \,
C^{(k)}(n \cdot p, \mu) \, \int_0^{\infty} {d \omega \over \omega- \bar n \cdot p}~
J^{(k)}\left({\mu^2 \over n \cdot p \, \omega},{\omega \over \bar n \cdot p}\right) \,
\phi_B^{(k)}(\omega,\mu)  \,, \nonumber \\
\widetilde{\Pi} &=& \tilde{f}_B(\mu) \, m_B \sum \limits_{k=\pm} \,
\widetilde{C}^{(k)}(n \cdot p, \mu) \, \int_0^{\infty} {d \omega \over \omega- \bar n \cdot p}~
\widetilde{J}^{(k)}\left({\mu^2 \over n \cdot p \, \omega},{\omega \over \bar n \cdot p}\right) \,
\phi_B^{(k)}(\omega,\mu)  \,,
\label{NLO factorization formula of correlator}
\end{eqnarray}
where the hard and hard-collinear functions at one loop read  \cite{Wang:2015vgv}
\begin{eqnarray}
C^{(+)}  &=& \tilde{C}^{(+)}=1, \qquad
C^{(-)} =\frac{\alpha_s \, C_F}{4 \, \pi}\, {1 \over \bar r} \,
\left [ {r \over \bar r} \, \ln r + 1 \right ]\,, \nonumber \\
\tilde{C}^{(-)} &=& 1 - \frac{\alpha_s \, C_F}{4 \, \pi}\,  \bigg [ 2 \, \ln^2 {\mu \over n \cdot p}
+ 5 \, \ln {\mu \over m_b} - \ln^2 r  - 2 \, {\rm Li_2} \left ( - {\bar r \over r} \right )
+ {2-r \over r-1} \, \ln r  +  {\pi^2 \over 12} + 5 \bigg ] \,,
\label{results of hard coefficients}
\end{eqnarray}
and
\begin{eqnarray}
J^{(+)}&=& {1 \over r} \, \tilde{J}^{(+)}
= \frac{\alpha_s \, C_F}{4 \, \pi} \, \left (1- { \bar n \cdot p \over \omega } \right ) \,
\ln \left (1- { \omega  \over \bar n \cdot p } \right ) \,,
\qquad  J^{(-)} = 1 \,, \nonumber \\
\tilde{J}^{(-)}&=& 1 + \frac{\alpha_s \, C_F}{4 \, \pi} \,
\bigg [ \ln^2 { \mu^2 \over  n \cdot p (\omega- \bar n \cdot p) }
- 2 \ln {\bar n \cdot p -\omega \over \bar n \cdot p } \, \ln { \mu^2 \over  n \cdot p (\omega- \bar n \cdot p) }
\,  \nonumber \\
&& - \ln^2 {\bar n \cdot p -\omega \over \bar n \cdot p }
- \left ( 1 +  {2 \bar n \cdot p \over \omega} \right )  \ln {\bar n \cdot p -\omega \over \bar n \cdot p }
-{\pi^2 \over 6} -1 \bigg ] \,.
\label{results of jet functions}
\end{eqnarray}
The hard coefficients presented in (\ref{results of hard coefficients}) are in compatible with the perturbative
matching coefficients of the weak current $\bar q \, \gamma_{\mu} \, b$ from QCD to soft-collinear effective
theory (SCET) \cite{Bauer:2000yr}, and the hard-collinear functions displayed in (\ref{results of jet functions}) coincide with
the corresponding jet functions  computed with the SCET Feynman rules \cite{DeFazio:2007hw}.

It is evident that there is no common value of $\mu$ that can avoid the parametrically large logarithms of order
$\ln(m_b / \Lambda)$ in the hard functions, the jet functions and the $B$-meson DA.
The summation of these large logarithms in the hard coefficient functions can be accomplished by solving the
following renormalization-group equations
\begin{eqnarray}
{d \over d \ln \mu} \tilde{C}^{(-)}(n \cdot p, \mu) &=&
\left [ - \Gamma_{\rm cusp}(\alpha_s) \ln { \mu \over n \cdot p} + \gamma(\alpha_s) \right ]
\tilde{C}^{(-)}(n \cdot p, \mu)\,, \nonumber \\
{d \over d \ln \mu} \, \tilde{f}_B(\mu) &=& \tilde{\gamma}(\alpha_s)\, \tilde{f}_B(\mu) \,,
\label{general RGE of tildeC1}
\end{eqnarray}
where the three-loop cusp anomalous dimension and the two-loop $\gamma$ ($\tilde{\gamma}$) are needed
to achieve the next-to-leading-logarithmic (NLL) resummation. Applying the standard procedure for the
construction of QCD sum rules leads to
\begin{eqnarray}
&& f_{\pi} \,\, e^{-m_{\pi}^2/(n \cdot p \, \omega_M)} \,\,\,
\left \{ \frac{n \cdot p} {m_B} \, f_{B \pi}^{+}(q^2)
\,, \,\,\,   f_{B \pi}^{0}(q^2)  \right \}  \,  \nonumber \\
&& = \left [U_2(\mu_{h2},\mu ) \, \tilde{f}_B(\mu_{h2}) \right ]
\,\,\, \int_0^{\omega_s} \,\, d \omega^{\prime} \,\, 
\, e^{-\omega^{\prime} / \omega_M}  \, \bigg [  \left [ U_1(n \cdot p,\mu_{h1},\mu ) \,
\widetilde{C}^{(-)}(n \cdot p, \mu_{h1})  \right ]
\, \phi_{B, \rm eff}^{-}(\omega^{\prime}, \mu)  \nonumber \\
&& \hspace{0.4 cm} + \, r \, \phi_{B, \rm eff}^{+}(\omega^{\prime}, \mu)
 \pm \,\,\, \frac{n \cdot p -m_B} {m_B}   \,\,\,
\left ( \phi_{B, \rm eff}^{+}(\omega^{\prime}, \mu)
+ C^{(-)}(n \cdot p, \mu) \, \phi_{B}^{-}(\omega^{\prime}, \mu) \right )  \bigg ] \,,
\label{NLO sum rules of form factors}
\end{eqnarray}
where the explicit expressions of $\phi_{B, \rm eff}^{\pm}(\omega^{\prime}, \mu)$
can be found in \cite{Wang:2015vgv}.

\begin{figure}
\centering
\includegraphics[width=10 cm,clip]{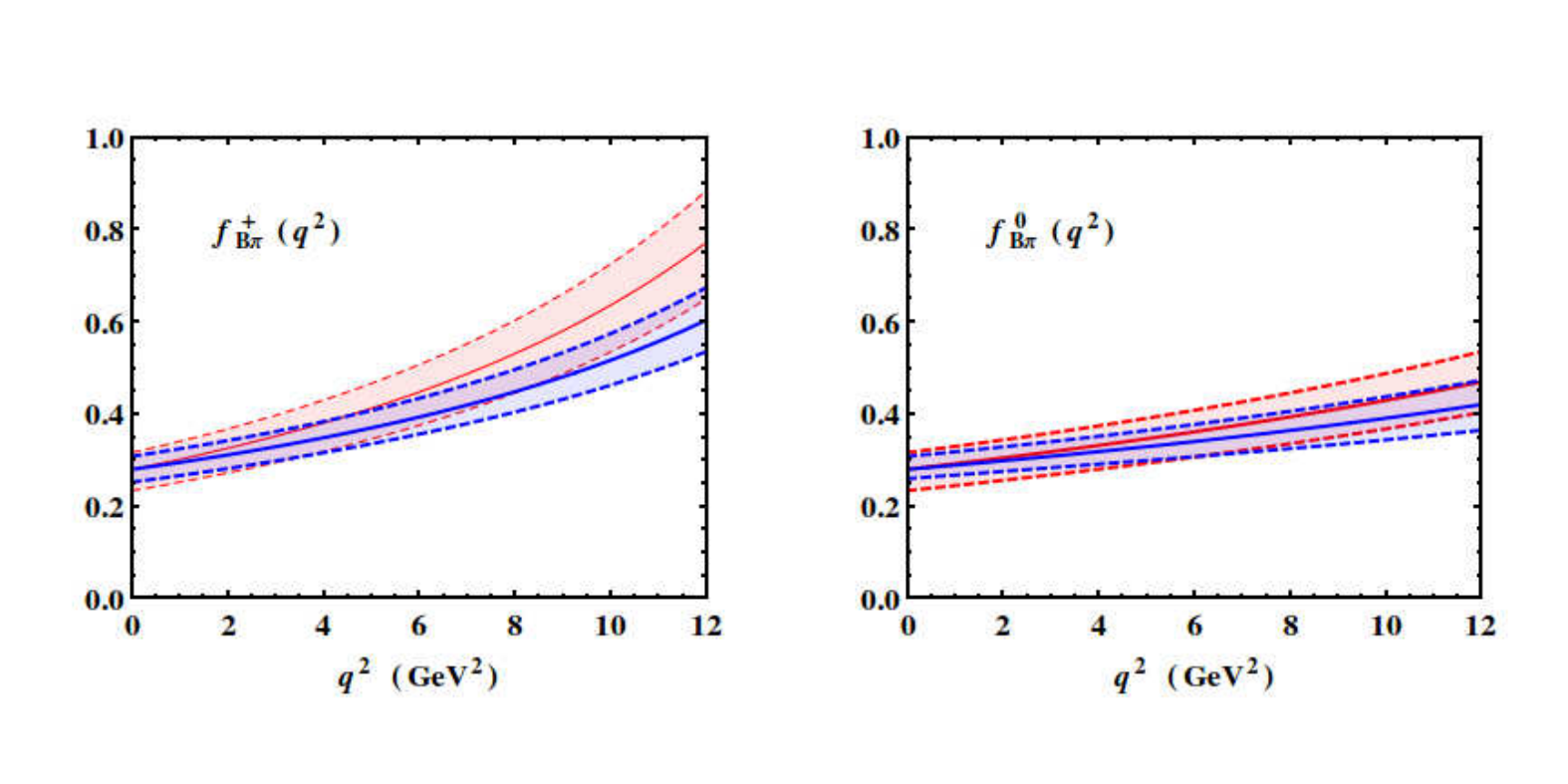}
\vspace*{0.1cm}
\caption{The momentum-transfer dependence of  the $B \to \pi$ form factors $f_{B \pi}^{+,0}(q^2)$
from the NLL resummation improved LCSR (\ref{NLO sum rules of form factors}) with the $B$-meson DA (pink curves)
and from the NLO QCD sum rules with the pion DA (blue curves). }
\label{final form factor shape}       
\end{figure}

Having at our disposal the NLL resummation improved LCSR presented in (\ref{NLO sum rules of form factors}),
it is straightforward to plot the $q^2$ dependence of the $B \to \pi$ form factors $f_{B \pi}^{+,0}(q^2)$
in figure \ref{final form factor shape} where the theoretical predictions from the sum rules with the pion DA
\cite{Khodjamirian:2011ub} are also shown for a comparison. The observed discrepancy of the $q^2$ shape for
the vector form factor $f_{B \pi}^{+}(q^2)$ predicted from the two distinct sum rules
could be attributed to the systematical uncertainties generated by the different parton-hadron duality ansatz
and to the yet unaccounted higher order/power corrections in the perturbative calculations of the corresponding
correlation functions.  Expressing the differential branching fraction of $B \to \pi \mu \nu_{\mu}$ in terms of the
form factor $f_{B \pi}^{+}(q^2)$
\begin{eqnarray}
\frac{d \Gamma}{d q^2} \, (B \to \pi \mu \nu_{\mu}) &=& \frac{G_F^2 |V_{ub}|^2}{24 \pi^3}
\, |\vec{p}_{\pi}|^3 \, \, |f_{B \pi}^{+}(q^2)|^2 \,.
\label{differential distribution formula: muon}
\end{eqnarray}
and employing the experimental measurements for the integrated  decay rate \cite{Lees:2012vv,Sibidanov:2013rkk}
lead to
\begin{eqnarray}
|V_{ub}|= \left(3.05^{+0.54}_{-0.38} |_{\rm th.} \pm 0.09 |_{\rm exp.}\right)  \times 10^{-3} \,,
\end{eqnarray}
in consistent with the exclusive determination of $|V_{ub}|$ from the leptonic
$B \to \tau \nu_{\tau}$ decay \cite{Kronenbitter:2015kls}.


\section{Conclusion}

To summarize, we demonstrated QCD factorization for the vacuum-to-$B$-meson correlation function
used in the construction of the $B \to \pi$ form factors at one-loop accuracy explicitly with
the method of regions and achieved the resummation of large logarithms in the short-distance functions
at NLL by solving the renormalization-group equations in the momentum space.
Phenomenological applications of the newly derived sum rules with the $B$-meson DA were also discussed in brief,
focusing on the extraction of exclusive $|V_{ub}|$ from the semi-leptonic $B \to \pi \mu \nu_{\mu}$ decay.

\label{intro}



%

\end{document}